\documentclass[aip,jap]{revtex4-1}
\usepackage{graphicx}
\usepackage{amsmath}
\usepackage{textcomp}

\begin{document}

\title{Use of THz Photoconductive Sources to Characterize Tunable Graphene RF Plasmonic Antennas}



\author{A. Cabellos}
\author{I. Llatser}
\author{E. Alarc\'on}
\affiliation{Nanonetworking Center in Catalonia, BarcelonaTech, Barcelona, Spain}
\author{A. Hsu}
\author{T. Palacios}
\affiliation{Massachusetts Institute of Technology, Cambridge, Massachusetts 02139,USA}
\date{\today}

\begin{abstract}
Graphene, owing to its ability to support plasmon polariton waves in the terahertz frequency range, enables the miniaturization and electrical tunability of antennas to allow wireless communications among nanosystems. One of the main challenges in the characterization and demonstration of graphene antennas is finding suitable terahertz sources to feed the antenna. This paper characterizes the performance of a graphene RF plasmonic antenna fed with a photoconductive source. The terahertz source is modeled and, by means of a full-wave EM solver, the radiated power as well as the tunable resonant frequency of the device is estimated with respect to material, laser illumination and antenna geometry parameters. The results show that with this setup, the antenna radiates terahertz pulses with an average power up to 1~\hbox{\textmu}W and shows promising electrical frequency tunability.
\end{abstract}


\maketitle 

\section{Introduction}

Graphene has recently attracted intense attention of the research community due to its extraordinary mechanical, electronic and optical properties~\cite{Geim2007}. One particularly promising research field is that of graphene-enabled wireless communications. The interconnection of nanosystems, i.e., integrated systems with a size of a few micrometers, has several applications, such as enabling wireless nanosensor networks~\cite{Akyildiz2010} or wireless communications in network-on-chip environments~\cite{Abadal2012b}.

However, wireless communications among nanosystems cannot be achieved by simply reducing the size of classical metallic antennas down to a few micrometers~\cite{Llatser2012c}. This approach presents several drawbacks, such as the low conductivity of nanoscale metallic structures and, especially, the very high resonant frequencies (up to the infrared and optical ranges) of micrometer-size antennas, which would result in a large channel attenuation and the difficulty of implementing transceivers operating at such a high frequency. 

For these reasons, using micrometer-size metallic antennas to implement wireless communications among nanosystems becomes very challenging. However, graphene offers a new approach to THz communications thanks to its ability to support the propagation of Surface-Plasmon Polariton (SPP) waves in the terahertz frequency band~\cite{Hanson2008a,PhysRevB.80.245435,citeulike:9922613}. Indeed, a graphene RF plasmonic antenna with lateral dimensions of just a few micrometers is predicted to resonate in the terahertz band~\cite{Jornet2010a,Llatser2011b}, at a frequency up to two orders of magnitude lower~\cite{Llatser2012c} and with a higher radiation efficiency with respect to typical THz metallic antennas~\cite{Tamagnone2012a, efficiency11}. Moreover, the electrical conductivity of graphene can be tuned by the application of an external electrostatic bias, which allows the design of tunable graphene antennas~\cite{Llatser2012a}. 

In the context of wireless communications among nanosystems, it is important to have a better understanding of graphene antennas by characterizing them towards its practical realization. One of the main challenges is that, because of its typical resonant frequency (1-10 THz), the antenna must be fed with a suitable terahertz source that can contact it with a reasonable efficiency. In recent years, significant research efforts have been devoted to push the limit of RF sources to higher frequencies, as well as to manufacture optical sources with longer emission wavelengths. Following the RF approach, resonance tunneling diodes~\cite{Eisele98,Brown91} and chains of frequency multipliers~\cite{Momeni11} provide compact terahertz sources, but they still show a poor efficiency and limited bandwidth. In the optical domain, Quantum-Cascade Lasers (QCL) have recently shown a significant progress, extending their operational frequency to 1~THz at cryogenic temperatures~\cite{Kohler02,Scalari09}. Additionally, optical down-conversion of ultrashort laser pulses by means of photoconductor materials has demonstrated a sustained increase in performance in the last decades~\cite{Auston84,Berry13}. In this approach, an ultrashort laser pulse illuminates the surface of a photoconductor generating photocarriers, which move under the influence of an external electrical bias field. The resulting photocurrent forms a picosecond pulse which, with the help of an antenna attached to the photoconductor converts into free-space terahertz radiation. 

Among the different approaches for terahertz sources, optical down-conversion is one of the best suited for the characterization and demonstration of graphene antennas for two main reasons. First, such terahertz sources have typically a very high impedance, in the order of several k$\Omega$, which lies in the same order of magnitude of the input impedance of graphene antennas~\cite{Tamagnone2012a}, thereby improving impedance matching between both devices. Second, ultrafast photoconductive antennas operate in pulsed mode, which is also the proposed fundamental mechanism for EM communications among nanosystems~\cite{Jornet2011}. In this paper, we first design a graphene RF plasmonic antenna fed with a photoconductive source and we characterize its radiated terahertz power by means of full-wave EM simulations. Our findings show that the terahertz signal radiated by the proposed device has a power in the \hbox{\textmu}W range, i.e., in the same order of magnitude as traditional photoconductive antennas, while having a size up to two orders of magnitude smaller~\cite{Llatser2012c}. Additionally we estimate the electrical frequency tunability of the antenna .

In what follows, the proposed device is described. Next, the photoconductive source as well as the antenna are modeled. Using these models and by means of full-wave EM solvers, the radiated terahertz power by a photoconductor-fed graphene antenna is estimated as a function of material, illumination and antenna geometry parameters.

\section{Device Description}

\begin{figure}
\begin{centering}
\includegraphics[width=110mm]{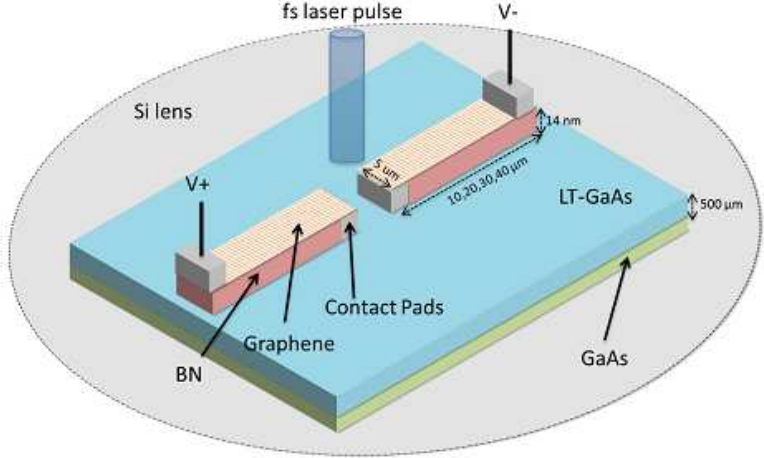} 
\end{centering}
\caption{Schematic representation of a graphene RF plasmonic antenna fed with a photoconductive material.}
\label{fig:sketch}
\end{figure}

Fig.~\ref{fig:sketch} depicts a schematic representation of the proposed graphene dipole antenna fed with a photoconductor. When a femtosecond laser pulse excites the biased semiconductor --Low-Temperature-grown GaAs (LT-GaAs)~\cite{Liu03}-- with a photon energy greater than its bandgap, electrons and holes are produced at the illumination point in the conduction and valence bands, respectively. The rapid changes of the density of the photocarriers and their acceleration to the applied voltage between the antenna edges produce a picosecond voltage pulse. 

This picosecond pulse excites Surface-Plasmon Polariton (SPP) waves at the interface between the graphene layer and the dielectric material. This SPP wave propagates along the graphene antenna producing a free-space terahertz radiation. The resonant frequency of the graphene antenna can be dynamically tuned with the application of an electrostatic voltage $V_{bias}$. As shown in Fig.~\ref{fig:sketch}, the dielectric material is boron nitride (h-BN) in order to achieve a high electron mobility~\cite{Dean10}. A high electron mobility is an important requirement for graphene antennas, given its strong dependence with the antenna efficiency, as further discussed in the results section. Finally, the silicon lens shown in Fig.~\ref{fig:sketch} is a common technique used to improve the directivity of the radiated signal~\cite{Tani97}.

\section{Photoconductor model}
\label{photoconductor}

In this section we present a model of the time-dependent voltage generated by a photoconductive source, based on a Drude-Lorentz model~\cite{Auston84}. The trapping time $\tau_c$ of the photocarriers in LT-GaAs is shorter than the recombination time of electrons and holes. Then, the charge density $n(t)$ can be obtained as

\begin{equation}
\frac{d n(t)}{dt} = - \frac{n(t)}{\tau_c} + G(t)
\end{equation}
where $G(t)$ is the generation rate of the carriers as a result of the laser pulse. The excitation time of the laser pulse is typically in the order of tens of fs. The velocity of the carriers $v(t)$ can be expressed as

\begin{equation}
\frac{d v(t)}{dt} = - \frac{v(t)}{\tau_s} + \frac{e}{m} E_{loc}
\end{equation}
where $\tau_s$ is the momentum relaxation time and $E_{loc}$ is the local field which can be expressed as $E_{loc}=E_{bias} - P_{sc}(t) / \eta \epsilon$, where $E_{bias}$ is the applied bias field, $\epsilon$ is the permittivity of the semiconductor material and $\eta$ is a geometric factor. Finally, the polarization $P_{sc}(t)$ caused by the separation of the electron and hole has the following expression:

\begin{equation}
\frac{d P_{sc}(t)}{dt} = - \frac{P_{sc}(t)}{\tau_r} + j(t)
\end{equation}
where $\tau_r$ is the recombination time of electrons and holes and $j(t)$ is the generated photocurrent density.

Based on this fundamental principles and under the assumption that the momentum relaxation time $\tau_s$ is small, \textit{Khiabani et al.}~\cite{Khiabani2013} developed an equivalent circuit model using lumped elements that provides accurate estimates --in agreement with published experimental data-- taking into account the underlying physical behavior. With this model, the time-dependent voltage at the antenna terminals $V$ can be expressed as:

\begin{equation}
V(t) = Z_a \cdot j(t) \beta \cdot V_c(t)
\label{eq:V}
\end{equation}
where $Z_a$ is the impedance of the graphene antenna (assumed as independent of the frequency), $V_c$ is the voltage at the antenna gap (derived from its equivalent circuit~\cite{Khiabani2013}) and $\beta$ is the active area in the semiconductor through which the photocurrent flows.

\begin{figure}
\begin{centering}
\includegraphics[width=100mm]{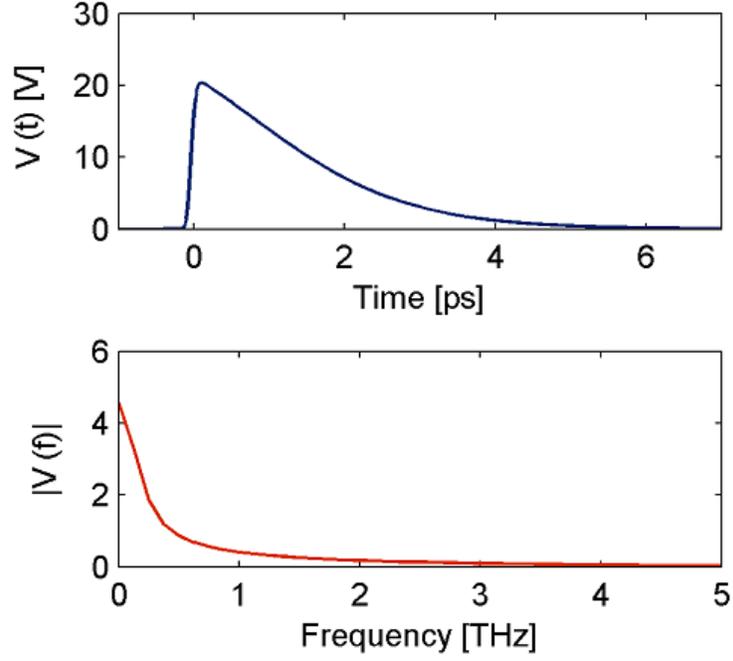} 
\end{centering}
\caption{Generated voltage by the photoconductive source $V$, in the time domain (above) and frequency domain (below).}
\label{fig:pulse}
\end{figure}

Fig.~\ref{fig:pulse} shows (using eq. \ref{eq:V}) the generated voltage pulse both in the time and frequency domain. This pulse excites SPP waves at the graphene antenna which result in free-space radiation. The main parameters of the photoconductive antenna are shown in Table~\ref{parameters}, which have been taken assuming realistic values from the literature~\cite{Khiabani2013}.

\begin{table}
\begin{center}
  \begin{tabular}{ l  c  c }
    \hline
    Parameter & Notation & Value \\ \hline
    Electron mobility for LT-GaAs & $\mu_{\textrm{GaAs}}$ & 200~cm$^2$V$^{-1}$s$^{-1}$ \\
	Average optical power & $P_{av}$ & 2~mW \\
	Laser frequency & $\nu_{opt}$ & 375~THz \\
	Laser repetition rate & $f_{rep}$ & 80~MHz \\
	Laser pulse duration & $\tau_l$ & 100~fs \\
	Carrier lifetime & $\tau_c$ & 1~ps \\
	Carrier recombination time & $\tau_r$ & 100~ps \\
    Bias voltage & $V_{bias}$ & 30~V \\
    Antenna gap length & $L$ & 10~\hbox{\textmu}m \\
    Antenna gap width & $W$ & 10~\hbox{\textmu}m \\
    Antenna resistance & $Z_a$ & 10~k$\Omega$ \\
    \hline
  \end{tabular}
\end{center}
\caption{Main parameters of the photoconductive antenna.}
\label{parameters}
\end{table}

\section{Antenna model}

The graphene antenna considered in this work is composed of a planar graphene dipole, over the same dielectric substrate as the photoconductive source. The choice of graphene for the implementation of this miniaturized antenna is twofold. First, graphene shows excellent conditions for the propagation of SPP waves at frequencies in the terahertz band~\cite{Gan2012,Nikitin2011}, that is, much lower frequencies than the SPP waves observed in noble metals, typically in the optical domain~\cite{Novotny2006}. Second, graphene allows to dynamically tune the antenna properties by means of an electrostatic bias~\cite{Llatser2012a}.

The radiation properties of graphene RF plasmonic antennas are mainly determined by the highly frequency-dependent character of the graphene electrical conductivity.  We consider a surface conductivity model for infinitely-large graphene sheets, which is valid for graphene structures with dimensions greater than a few hundred nanometers and it has been both theoretically~\cite{Falkovsky2007a} and experimentally validated~\cite{Rouhi2012,Horng2011}. The surface conductivity of an infinite graphene film can be calculated by means of the Kubo formalism~\cite{Falkovsky2007a,Hanson2008a}. Within the frequency region of interest (below 5~THz), the surface conductivity may be approximated by its intraband contribution~\cite{Llatser2012a}. Following the random-phase approximation, the intraband conductivity of graphene is expressed with a Drude model:

\begin{equation}
\sigma\left(\omega\right)=\frac{2e^{2}}{\pi\hbar}\frac{k_{B}T}{\hbar}\ln\left[2\cosh\left[\frac{E_F}{2k_{B}T}\right]\right]\frac{i}{\omega+i\tau^{-1}},\label{eq:sigma_graphene}
\end{equation}
where $\tau$ is the relaxation time, $T$ is the temperature and $E_F$ is the chemical potential. The relaxation time is obtained as $\tau = \hbar \mu_{\textrm{g}} \sqrt{\pi  n} / e v_p$, where $\mu_{\textrm{g}}$ is the electron mobility in graphene, which depends on the quality of graphene and the dielectric substrate, amongst others, $n = 4 \pi (E_F / h v_p)^2$ the number of electrons and $v_F = c/300$ the Fermi velocity.

In order to study the radiation properties of these antennas, we consider a simple Fabry-Perot resonator model. A graphene planar dipole placed on an air-dielectric interface supports transverse-magnetic (TM) SPP waves with a dispersion relation given by~\cite{Jablan2009}

\begin{equation}
\frac{1}{\sqrt{k_{\mathrm{SPP}}^{2}-\frac{\omega^{2}}{c^{2}}}}+\frac{\varepsilon}{\sqrt{k_{\mathrm{SPP}}^{2}-\varepsilon\frac{\omega^{2}}{c^{2}}}}=-\mathrm{i}\frac{\sigma\left(\omega\right)}{\omega\varepsilon_{0}},
\label{eq:DR}
\end{equation}
where $k_{\mathrm{SPP}}$ and $\omega$ are the wavevector and frequency of SPP waves, and $\varepsilon$ is the dielectric constant of the substrate. We observe that the graphene conductivity has a crucial impact in the dispersion relation and, ultimately, in the relationship between wavelength and frequency of SPP waves in graphene antennas.

The dispersion relation allows the derivation of the antenna resonant frequency by means of the resonance condition:

\begin{equation}
L = m\frac{\lambda_{\mathrm{spp}}}{2} = m\frac{\pi}{k_{\mathrm{SPP}}},
\label{eq:res_condition}
\end{equation}
where $L$ is the antenna length, $m$ is an integer corresponding to the resonance order and $\lambda_{\mathrm{spp}}$ is the SPP wavelength. Solving the dispersion relation \eqref{eq:DR} with the first-order ($m=1$) resonance condition \eqref{eq:res_condition} allows obtaining the resonant frequency of the graphene antenna for a given antenna length.

\begin{figure}
\begin{centering}
\includegraphics[width=100mm]{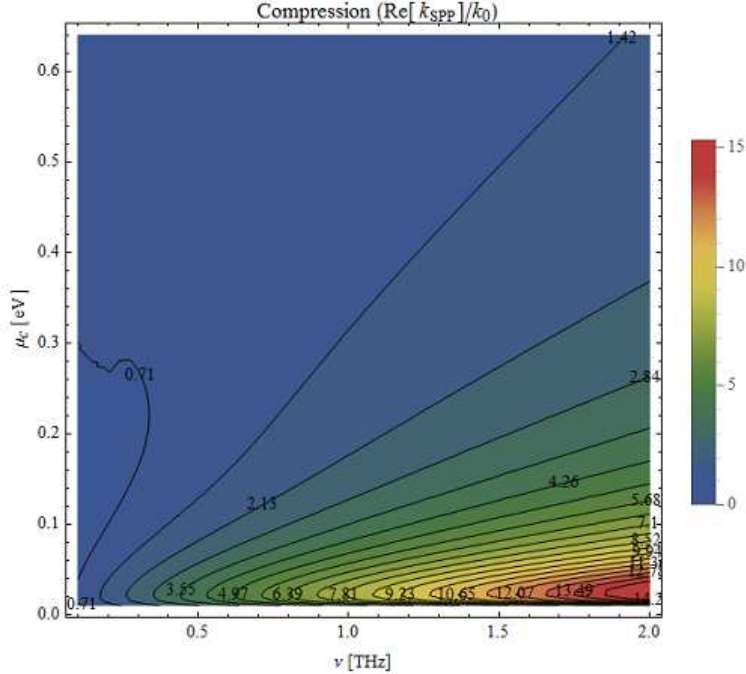} 
\end{centering}
\caption{Plasmon compression factor $K$ in a graphene RF plasmonic antenna as a function of the frequency and chemical potential.}
\label{fig:compression}
\end{figure}

Two important performance metrics that allow to quantify the radiation properties of plasmonic antennas are the \textit{plasmon compression factor} and the \textit{plasmon propagation length}. The plasmon compression factor $K$ is defined as the ratio between the free-space wavelength and the plasmon wavelength: $K = \lambda_0 / \lambda_{\textrm{spp}}$ or, equivalently, the ratio between the real part of the SPP wavevector and the free-space wavevector  $K = \mathrm{Re}[k_{\mathrm{SPP}}] / k_0$. Moreover, this parameter determines the size difference between the graphene RF plasmonic antenna and a metallic non-plasmonic antenna resonating at the same frequency, i.e., $K = L_0/L$, where $L_0$ is the metallic antenna length and $L$ that of the graphene antenna. Figure~\ref{fig:compression} shows the plasmon compression factor as a function of the frequency and the chemical potential. A graphene layer with an electron mobility of 10000~cm$^2/$Vs over a dielectric substrate with dielectric constant $\varepsilon=4$ are considered. We observe that a desired high compression factor is obtained for high plasmon frequencies and a low (but non-zero) chemical potential.

The plasmon propagation length $L_{\textrm{spp}}$ is defined as the distance for the SPP intensity to decay by a factor of $1/e$ and it has a direct impact on the radiation efficiency of the graphene antenna. In the scenario under consideration, there is a linear relationship between the antenna radiation efficiency and the plasmon propagation length in the graphene sample; therefore, a large $L_{\textrm{spp}}$ will result in a high transmitted power by the graphene antenna.

\begin{figure}
\begin{centering}
\includegraphics[width=100mm]{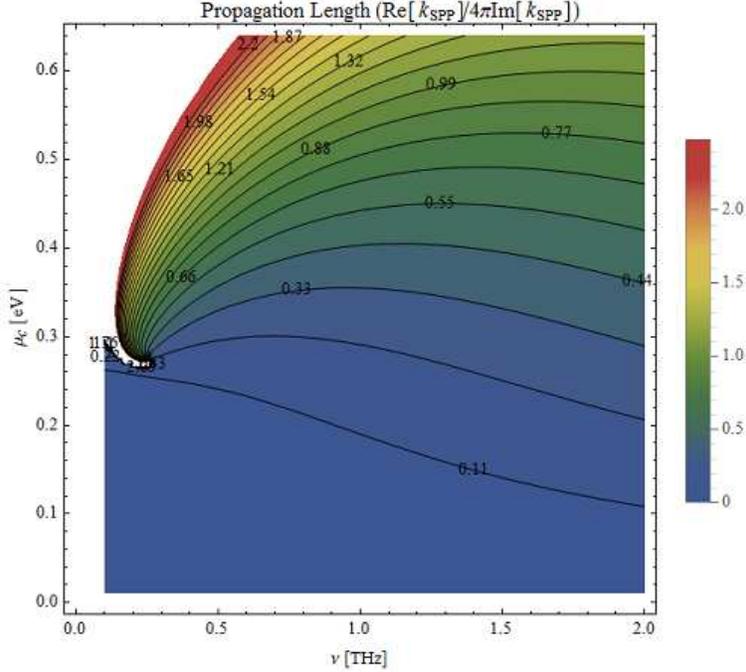} 
\end{centering}
\caption{Plasmon propagation length $L_{\textrm{spp}}$ in a graphene RF plasmonic antenna as a function of the frequency and chemical potential.}
\label{fig:length}
\end{figure}

The plasmon propagation length is usually expressed as a function of the plasmon wavelength, yielding the expression $L_{\textrm{spp}} / \lambda_{\textrm{spp}} = \textrm{Re}[k_{\mathrm{SPP}}] / 4 \pi \textrm{Im}[k_{\mathrm{SPP}}]$. Fig.~\ref{fig:length} shows the plasmon propagation length in units of the corresponding SPP wavelength, with the same parameters as in the previous case. Opposite to the case of the plasmon compression factor, the plasmon has a longer propagation length (of above one plasmon wavelength) in the low frequency and high chemical potential potential region. Therefore, there exists a trade-off between the plasmon compression factor and its propagation length: graphene antennas operating at a high frequency and low chemical potential will allow a higher miniaturization with respect to metallic antennas, whereas graphene antennas working a low frequency and high chemical potential will exhibit lower losses and, in consequence, a higher radiation efficiency.

\section{Results}

We show next some results of the radiated power by the photoconductive source-fed graphene antenna, as a function of several parameters from both the antenna and the photoconductive source. We obtain the power of the terahertz pulse generated by the photoconductive source as $P = V^2/Z_a$, where $V$ is obtained by solving the model previously presented in Section~\ref{photoconductor}. Then, the power radiated by the graphene antenna is obtained by taking into account the two main impairments introduced by the graphene antenna. First, the impedance mismatch loss between the photoconductive source and the graphene antenna: $M_L = 1-\rho^2$, where $\rho$ is the reflection coefficient $\rho=(Z_a-Z_s)/(Z_a+Z_s)$. Second, the antenna radiation efficiency $\varepsilon_R$, which is calculated numerically by means of the full-wave electromagnetic simulator FEKO. The power radiated by the antenna $P_{rad}$ is then calculated as

\begin{equation}
P_{rad} = M_L \cdot \varepsilon_R \cdot P
\label{rad_power}
\end{equation}

\begin{figure}
\begin{centering}
\includegraphics[width=100mm]{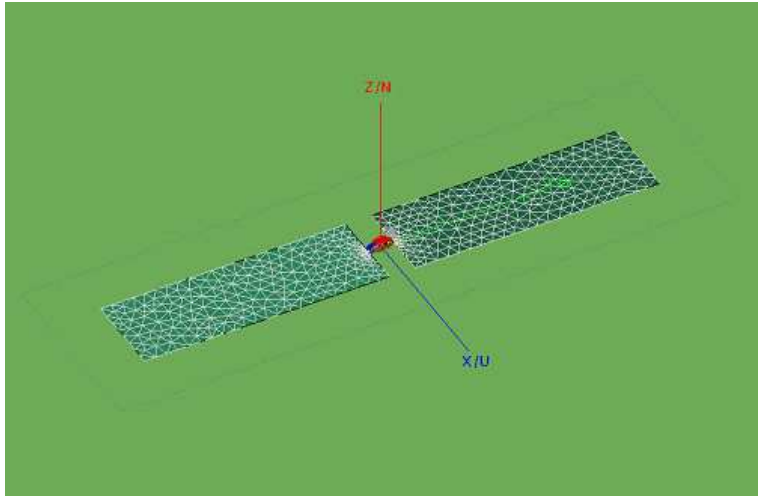} 
\end{centering}
\caption{Schematic model of the graphene RF plasmonic antenna.}
\label{fig:antenna}
\end{figure}

The graphene antennas are modeled as a center-fed graphene planar dipole with a width of 5~\hbox{\textmu}m and different lengths, over an infinite LT-GaAs substrate with a thickness of 500~\hbox{\textmu}m and a dielectric constant $\varepsilon = 12.9$. The electrically-thin h-BN substrate (with a thickness of 14~nm, less than $\lambda$/1000) has a negligible impact in the simulation results. Figure~\ref{fig:antenna} shows a diagram of the antenna model used in the simulations.

For this analysis, we consider the parameters for the photoconductive source in Table~\ref{parameters}, unless otherwise specified. The impedance of the photoconductive source is obtained from the source conductance~\cite{Khiabani2011} as $Z_s(t) = 1/G_s(t)$. Considering a photoconductive source with the described parameters, the source impedance has a minimum value of 2.24~k$\Omega$, consistently with typical values for photoconductors. The antenna input impedance is obtained by numerical simulation and it ranges from 0.73~k$\Omega$ to 3.15~k$\Omega$, presenting values up to two orders of magnitude higher than comparable metallic antennas.

\begin{figure}
\begin{centering}
\includegraphics[width=120mm]{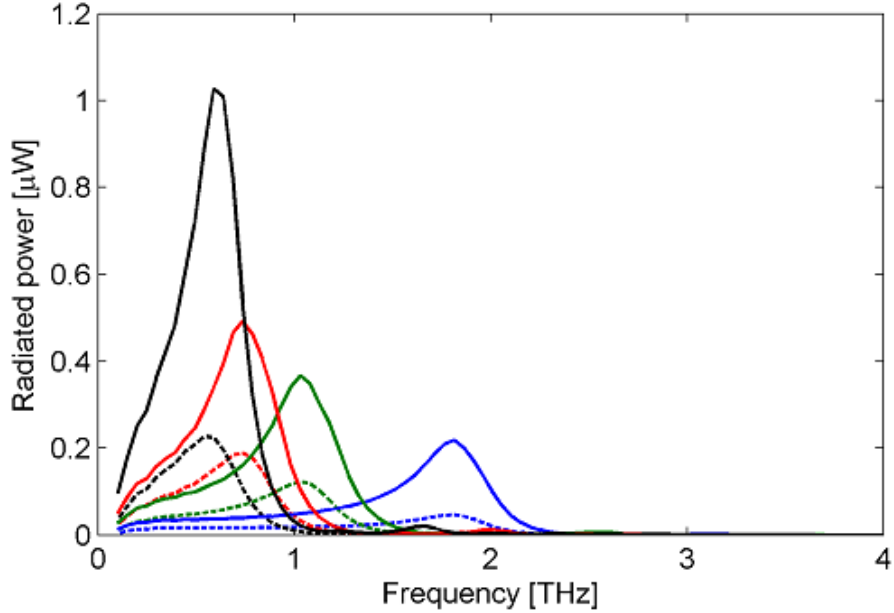} 
\end{centering}
\caption{Radiated power as a function of frequency for graphene antennas with lengths $L=10$~\hbox{\textmu}m (blue), 20~\hbox{\textmu}m (green), 30~\hbox{\textmu}m (red) and 40~\hbox{\textmu}m (black), from right to left, fed by a photoconductive source. The electron mobilities are $\mu_{\textrm{g}} = 20000$~cm$^2$/Vs (solid lines) and $\mu_{\textrm{g}} = 10000$~cm$^2$/Vs (dashed lines).}
\label{fig:power}
\end{figure}

Fig.~\ref{fig:power} shows the average radiated power as a function of frequency for graphene antennas with lengths of 10, 20, 30 and 40~\hbox{\textmu}m (from right to left) and width of 5~\hbox{\textmu}m, fed by the previously described photoconductive source. We compare two values for the electron mobility of the graphene layer: $\mu_{\textrm{g}} = 20000$~cm$^2$/Vs (solid lines) and $\mu_{\textrm{g}} = 10000$~cm$^2$/Vs (dashed lines). Electron mobilities of graphene over boron nitride of up to 40000~cm$^2$/Vs have been experimentally observed at room temperature~\cite{Dean10}; therefore, the selected values are conservative estimates. The chemical potential is $E_F = 0.4$~eV. We observe that the spectrum of the radiated signal presents clear resonances in the \hbox{\textmu}W range, due to the highly frequency-selective behavior of the graphene dipole antenna. Similarly to the behavior observed in metallic antennas, the resonant frequency of a graphene antenna decreases as its length increases, although following a different scaling trend~\cite{Llatser2012c}. As expected, the radiated power is reduced in the case of the lower electron mobility, mainly due to the lower plasmon propagation length and the associated reduced radiation efficiency of the graphene antenna.

As shown in Eq.~\eqref{rad_power}, the power radiated by the graphene antenna depends on three main factors: the output power of the photoconductive source, the impedance mismatch loss and the antenna radiation efficiency. The results in the considered scenario indicate that, on the one hand, longer antennas resonate at lower frequencies where the power density of the pulses generated by the photoconductive source is mostly concentrated (see Fig.~\ref{fig:pulse}), yielding a higher output power. On the other hand, longer graphene antennas also have a higher impedance mismatch and lower radiation efficiency than their shorter counterparts, which reduces the final radiated power. Overall, the higher output power of photoconductive sources in lower frequencies dominates and, as shown in Fig.~\ref{fig:power}, the radiated power by graphene antennas increases with the antenna length, reaching a maximum of 1~\hbox{\textmu}W for an antenna length of 40~\hbox{\textmu}m.

\begin{figure}
\begin{centering}
\includegraphics[width=110mm]{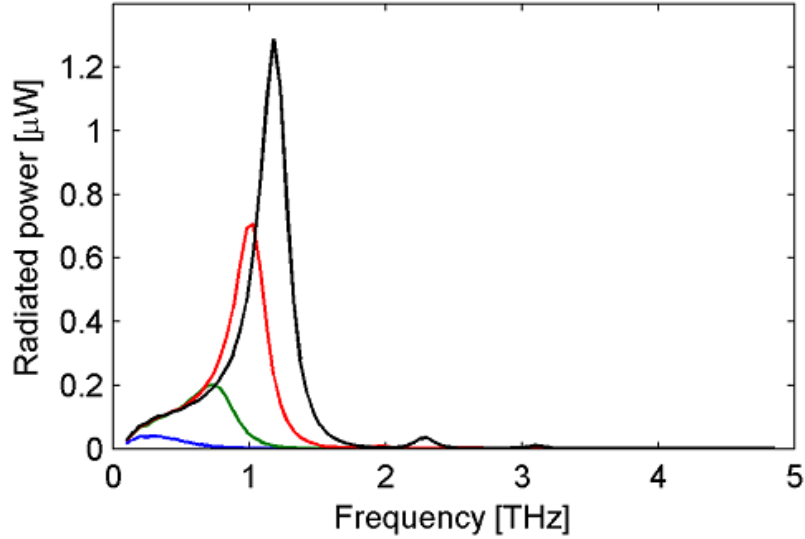} 
\end{centering}
\caption{Radiated power as a function of frequency for graphene antennas with dimensions 30x5~\hbox{\textmu}m, electron mobility $\mu_{\textrm{g}} = 10000$~cm$^2$/Vs and chemical potentials $E_F=0.2$~eV (blue), 0.4~eV (green), 0.7~eV (red) and 1~eV (black), from left to right, fed by a photoconductive source.}
\label{fig:power_Ef}
\end{figure}

We study next how applying an electrostatic bias voltage to the graphene layer influences the tunable resonant frequency of the antenna as well as the radiated power. This bias voltage causes a variation in the chemical potential of the graphene layer, which results in a higher conductivity value and, in consequence, the radiation efficiency of the graphene antenna is greatly enhanced. At the same time, the impedance mismatch is slightly increased and the resonant frequency moves to higher values, where the photoconductive source has a lower output power. Overall, the increase in the antenna radiation efficiency dominates and increasing the electrostatic bias applied to the graphene layer yields a higher radiated power. This is shown in Fig.~\ref{fig:power_Ef}, which plots the radiated power by graphene antennas with a fixed length of 30~\hbox{\textmu}m and width of 5~\hbox{\textmu}m, electron mobility $\mu_{\textrm{g}} = 10000$~cm$^2$/Vs and an applied chemical potential with values of 0.2, 0.4, 0.7 and 1~eV, from left to right. A radiated power of up to 1.3~\hbox{\textmu}W is obtained with a chemical potential of 1~eV.

\begin{figure}
\begin{centering}
\includegraphics[width=120mm]{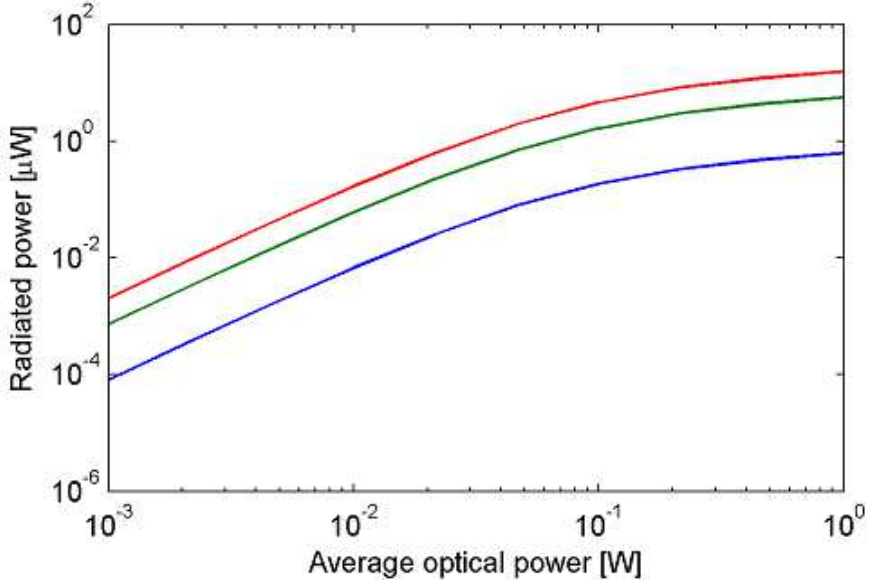} 
\end{centering}
\caption{Maximum radiated power by the graphene antenna as a function of the average optical power in the photoconductive source. The electron mobility is $\mu_{\textrm{g}} = 10000$~cm$^2$/Vs and the bias voltage is set to $V_{bias}=10$, 30 and 50~V (bottom to top).}
\label{fig:opt_power}
\end{figure}

In Fig.~\ref{fig:opt_power}, we observe how the power radiated by the graphene antenna scales with respect to the average optical power fed into the photoconductive source. The antenna length and width are of 30 and 5~\hbox{\textmu}m, respectively, the electron mobility $\mu_{\textrm{g}} = 10000$~cm$^2$/Vs and the chemical potential $E_F = 0.4$~eV. The radiated power scales linearly with respect to the power of the laser input up to an optical power of around 100~mW, when the radiated power begins to saturate. The bias voltage is applied to the photoconductive source along the antenna gap length $L$. Increasing the bias voltage (set to 10~V, 30~V and 50~V, from bottom to top) also enhances the power radiated by the graphene antenna.

\begin{figure}
\begin{centering}
\includegraphics[width=120mm]{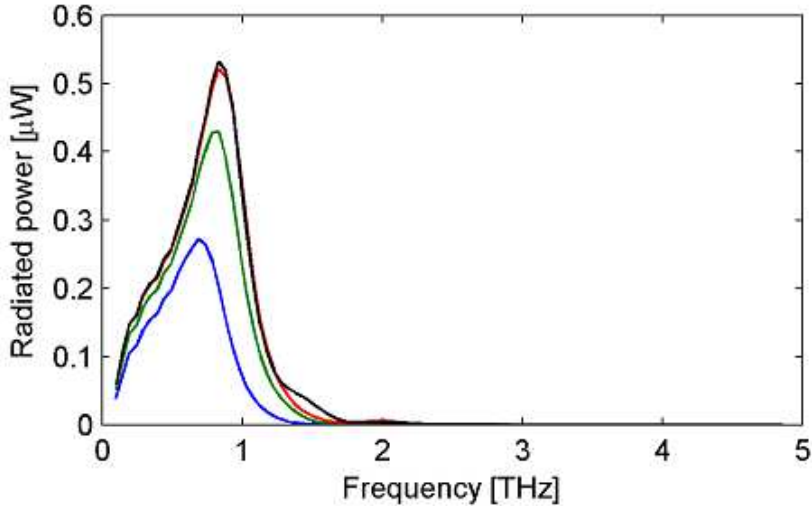} 
\end{centering}
\caption{Radiated power as a function of frequency for graphene bowtie antennas with a length $L=30$~\hbox{\textmu}m and widths $W=5$, 10, 15 and 20~\hbox{\textmu}m (bottom to top), fed by a photoconductive source. The electron mobility is $\mu_{\textrm{g}} = 10000$~cm$^2$/Vs.}
\label{fig:power_bowtie}
\end{figure}

Finally, we consider a different antenna geometry that is commonly used in photoconductive antennas due to its wideband behavior, the bowtie antenna. Fig.~\ref{fig:power_bowtie} shows the average power radiated by graphene bowtie antennas with a fixed length $L = 30$~\hbox{\textmu}m, electron mobility $\mu_{\textrm{g}} = 10000$~cm$^2$/Vs and chemical potential $E_F = 0.4$~eV, and different widths of 5, 10, 15 and 20~\hbox{\textmu}m, from bottom to top. We observe that the radiated power increases with the antenna width, as expected due to its larger area, whereas the antenna resonant frequency is relatively independent of its width. Comparing the radiated power of two graphene antennas with the same configuration but different geometry, one a bowtie antenna with a width of 10~\hbox{\textmu}m (Fig.~\ref{fig:power_bowtie}) and the other a planar dipole with a width of 5~\hbox{\textmu}m and a length of 20~\hbox{\textmu}m (Fig.~\ref{fig:power}), which have approximately the same area, we observe that the radiated power by the bowtie antenna is approximately twice that of the dipole antenna. This result confirms that, due to its wideband nature, the bowtie antenna is capable to radiate a higher fraction of the power generated by the photoconductive source and it is therefore better suited to maximize the output power of a photoconductive antenna.

\section{Conclusions}
\label{conclusions}

This paper characterizes by means of a full-wave EM solver electrically tunable graphene antennas fed with photoconductive terahertz source, showing a radiated power in the order of 1~\hbox{\textmu}W. The characterization of graphene antennas presents two main challenges. First, graphene antennas have a remarkable higher impedance (in the order of $k\Omega$) when compared to metallic antennas with an impedance of tens of ohms. However, in this context, due to the high impedance of photoconductive sources, graphene antennas can operate closer to impedance matching than metallic antennas. Second, graphene antennas have a lower radiation efficiency than metallic antennas due to the low electrical conductivity of graphene compared to a metal. This effect can be mitigated by using h-BN as a substrate for the graphene antenna, which increases the electron mobility of the graphene sample. Our results show that both effects compensate each other and graphene antennas radiate terahertz pulses with a power in the same order of magnitude than traditional metallic photoconductive antennas. Furthermore, graphene antennas have a size up to two orders of magnitude smaller than metallic antennas radiating at the same frequency, and they can be dynamically tuned by the application of an external bias voltage. These two unique features of graphene antennas make them a very promising candidate to implement wireless communications among nanosystems.

\begin{acknowledgments}
We would like to thank A.~Frenzel and Prof.~N.~Gedik for their valuable suggestions that greatly improved this work. This work has been partially supported by the FPU grant of the Spanish Ministry of Education.
\end{acknowledgments}



%

\end{document}